# Maximum Spherical Mean Value (mSMV) Filtering for Whole Brain Quantitative Susceptibility Mapping

Alexandra G. Roberts,[1,2] Dominick J. Romano,[2,3] Mert Şişman,[1,2] Alexey V. Dimov,[2] Pascal Spincemaille,[2] Thanh D. Nguyen,[2] Ilhami Kovanlikaya,[2] Susan A. Gauthier[2], Yi Wang.[1,2,3]

[1]Department of Electrical and Computer Engineering, Cornell University, Ithaca NY, USA

[2]Department of Radiology, Weill Cornell Medicine, New York, NY, USA

[3]Meinig School of Biomedical Engineering, Cornell University, Ithaca NY, USA

**Corresponding author**: Yi Wang, PhD

407 E 61st St, Ste RR 118

New York, NY 10021, USA

Email: yiwang@med.cornell.edu

## ABSTRACT

**Purpose:** To develop a tissue field filtering algorithm, called maximum Spherical Mean Value (mSMV), for reducing shadow artifacts in quantitative susceptibility mapping (QSM) of the brain without requiring brain tissue erosion.

**Theory and Methods:** Residual background field is a major source of shadow artifacts in QSM. The mSMV algorithm filters large field magnitude values near the border, where the maximum value of the harmonic background field is located. The effectiveness of mSMV for artifact removal was evaluated by comparing with existing QSM algorithms in numerical brain simulation as well as using in vivo human data acquired from 11 healthy volunteers and 93 patients.

**Results:** Numerical simulation showed that mSMV reduces shadow artifacts and improves QSM accuracy. Better shadow reduction, as demonstrated by lower QSM variation in the gray matter and higher QSM image quality score, was also observed in healthy subjects and in patients with hemorrhages, stroke and multiple sclerosis.

**Conclusion:** The mSMV algorithm allows QSM maps that are substantially equivalent to those obtained using SMV-filtered dipole inversion without eroding the volume of interest.

# INTRODUCTION

Quantitative susceptibility mapping (QSM) is an MRI method to calculate the tissue magnetic susceptibility from the tissue field (also referred to as the local field, as opposed to the background field generated by air) measured using complex gradient echo (GRE) images.[1] In QSM, the tissue field is a convolution of the tissue susceptibility distribution with the dipole kernel.[2-4] The inverse problem of computing susceptibility from the tissue field by means of dipole deconvolution is ill-posed and the partial differential equation corresponding to dipole deconvolution in the image space yields a wave propagator as a fundamental solution.[5,6] Therefore, dipole-incompatible tissue field components unexplained by the dipole convolution model cause streaking or shadow artifacts.[6-8] Bayesian inference QSM approaches suppress streaking artifacts using total variation (TV) regularization with spatial weighting as in Morphology Enabled Dipole Inversion (MEDI)[9] or without as in Fast Nonlinear Susceptibility Inversion (FANSI).[10] However, spatially slow-varying shadowing artifacts persist and methods such as Multi-Scale Dipole Inversion (MSDI)[11] reduce their influence on QSM. Shadows also persist near features with large susceptibility gradients such as hemorrhages. Dipole inversion algorithms such as STreaking Artifact Reduction (STAR)[12] QSM and robust masking and artifact reduction (QSMxT)[13] have been developed to suppress these artifacts.

Single reconstructions reducing shadow artifacts during QSM reconstruction[14,15] typically utilize the Spherical Mean Value (SMV) property of the harmonic background field. Most commonly, SMV-filtering was applied during background field removal, as in the Sophisticated Harmonic Artifact Reduction for Phase data (SHARP)[16] algorithm or its variations including Variable Sophisticated Harmonic Artifact Reduction (VSHARP),[17] Regularized Sophisticated Harmonic Artifact Reduction for Phase data (RESHARP),[18] and Improved Region-Adaptive Sophisticated

Harmonic Artifact Reduction (iRSHARP),[19] and iterative Spherical Mean Value (iSMV)[20] background field removal. These algorithms require optional or mandatory erosion of the brain mask as the SMV operation cannot be applied at the brain boundary where the filter kernel extends beyond the ROI, a limitation referred to as "kernel overlap". Other background field removal methods like the Laplacian Boundary Value (LBV) algorithm perform brain erosion to obtain high quality estimates of the total field at the boundary.[21] SMV-filtering can also be applied to the data term in iterative dipole inversion algorithms such as MEDI, which requires brain mask erosion. While these methods effectively remove QSM artifacts, brain mask erosion can lead to misdetection of pathologies including hemorrhages, microbleeds, and abnormal iron accumulation near the brain boundary.[22-24] Dipole inversion-based techniques that use SMV-filtered field in the data fidelity term to reduce shadow artifacts introduce erosion.[13,25] Other methods reduce shadow artifacts with smoothing regularization terms but result in underestimation in regions-of-interest (ROIs).[26,27] Here, an algorithm based on the maximum value property of harmonic functions is proposed to perform SMV-filtered dipole inversion while preserving the brain mask. This method is known as maximum Spherical Mean Value, or mSMV.

## THEORY

### Harmonic functions

Let $M$ be the mask where a susceptibility map is desired. Denote by $\partial M_0$ the boundary of $M(\boldsymbol{r})$ and $M_0$ the interior region, such that $M(\boldsymbol{r}) = \partial M_0 \cup M_0$. Define the total field $b(\boldsymbol{r})$ as a sum of the background field $b_b(\boldsymbol{r})$ and the tissue field $b_t(\boldsymbol{r})$,

$$b(\boldsymbol{r}) = b_b(\boldsymbol{r}) + b_t(\boldsymbol{r}) \tag{1}$$

The background field $b_b(\boldsymbol{r})$ is a harmonic function and satisfies the SMV property,[28] the basis of many of the methods discussed in this work. It states for point $\boldsymbol{c}$ within $M_0$ and spherical region $S(\boldsymbol{c})$ with volume $|S(\boldsymbol{c})|$, centered around $\boldsymbol{c}$ and completely within $M$, the background field satisfies:

$$\frac{1}{|S(\boldsymbol{c})|}\int_{S(\boldsymbol{c})} b_b(\boldsymbol{r})\, d\boldsymbol{r} = b_b(\boldsymbol{c}) \tag{2}$$

A harmonic function also satisfies the maximum value property - its maximum value over the region $M$ is always found at the boundary $\partial M_0$:[29]

$$\underset{\boldsymbol{r}\in M}{\operatorname{argmax}}\, b_b(\boldsymbol{r}) \in \partial M_0 \tag{3}$$

### Maximum SMV (mSMV) algorithm

The mSMV algorithm is applied to the local tissue field, denoted $\hat{b}_L(\boldsymbol{r})$ from conventional background field removal algorithms such as Projection onto Dipole Fields (PDF),[30] VSHARP,[17] or LBV.[21] It is known that background field removal methods underperform at the brain boundary, resulting in non-negligible residual background field.[31] By the maximum (and minimum) value property, and by the magnitude difference separating the tissue and residual background fields, mSMV assumes residual background field can be detected from the largest value of the estimated tissue field magnitude near the mask boundary. Exclusion of large sources in the tissue field such as veins or hemorrhages is addressed with an $R_2^*$ mask, discussed further in later paragraphs.

First, an initial SMV-filtering operation is performed,

$$b_L^0(\boldsymbol{r}) = \hat{b}_L(\boldsymbol{r}) - \left(S_{r_1} M \hat{b}_L\right)(\boldsymbol{r}) \tag{4}$$

Where the SMV operator $S_R$ for a given radius $R$ acting on a field $b$ is

$$(S_R b)(\boldsymbol{r}) = \mathcal{F}^{-1}\{\mathcal{F}\{b(\boldsymbol{r})\}\kappa_R\} \tag{5}$$

Here $\mathcal{F}$ denotes the Fourier transform, and $\kappa_R$ the Fourier transform of the spherical kernel with radius $R$. Let $\partial M_{r_1}$ be the mask of voxels within distance $r_1$ or less from the mask boundary $\partial M_0$. Voxels in this region are removed from the brain mask $M$ because their corresponding SMV kernel $\kappa_{r_1}$ overlaps with voxels outside $M$ (such as skull or air). Including these yields artifacts in the estimated susceptibility map.

In mSMV, it is assumed that $b_L^0(\boldsymbol{r})$ does not contain residual background field within $M_{r_1} = M \setminus \partial M_{r_1}$, i.e., for voxels further than a distance $r_1$ from the mask boundary. Inside $\partial M_{r_1}$, i.e., for voxels closer than a distance $r_1$ from the mask boundary, the aim is to identify voxels with residual background field contribution by selecting those where the absolute value of the estimated filtered field is larger than a threshold $t$. This threshold was empirically determined by taking the maximum value of the SMV-filtered local field magnitude as the kernel radius approaches 0, or

$$t = \lim_{r' \to 0} \max_{\boldsymbol{r} \in M} \left| b_L^0(\boldsymbol{r}) - (S_{r'} b_L^0)(\boldsymbol{r}) \right| \tag{6}$$

The kernel radius limit is based on minimizing error due to kernel overlap, or the inclusion of voxels outside of the brain mask in the mean value calculation. This is because, in continuous space, the kernel radius can become arbitrarily small and all background field contributions to $b_L^0(\boldsymbol{r})$ are removed. Then, the corresponding maximum of remaining tissue field is a reasonable estimate of this threshold.[32] In practice, discretization errors in the SMV operation become large for small radii, not allowing this threshold below a certain value produces reasonable results:

$$t = \max\left( t_{min}, \lim_{r' \to 0} \max_{\boldsymbol{r} \in M} \left| b_L^0(\boldsymbol{r}) - (S_{r'} b_L^0)(\boldsymbol{r}) \right| \right) \tag{7}$$

where $t_{min} \sim \mathcal{O}(10^{-1})\ Hz$ at $3\ T$ and is scaled accordingly for other field strengths. Masks obtained using this threshold will likely contain cortical veins, which have high field values indistinguishable from voxels at the brain edge that still contain background field. Therefore, a vein segmentation mask $M_v$ was constructed by applying a Frangi filter[33] to the $R_2^*$ map computed from the same gradient echo data, followed by thresholding. Voxels inside this mask are excluded from further processing (see below).

Next, for a given maximum number of iterations $i_{max}$ and $i = 1, \dots, i_{max}$, define the mask $M_{b,v}^i(\boldsymbol{r})$ as:

$$M_{b,v}^i(\boldsymbol{r}) = \begin{cases} 1 & \left| b_L^{i-1}(\boldsymbol{r}) \right| > t,\ \boldsymbol{r} \in \partial M_{r_1},\ \boldsymbol{r} \notin M_v \\ 0 & \text{else} \end{cases} \tag{8}$$

With this mask, define $b_L^i(\boldsymbol{r})$ as

$$b_L^i(\boldsymbol{r}) = b_L^{i-1}(\boldsymbol{r}) - \left(S_{r_2} M_{b,v}^i b_L^{i-1}\right)(\boldsymbol{r}) \tag{9}$$

with a modified SMV radius $r_2$ corresponding to the smallest sphere such that $(S_{r_2} b)(\boldsymbol{r}) \neq \delta(\boldsymbol{r})$ for the given resolution. Practically, this is defined as

$$r_2 = \frac{1}{2}\min\left(\Delta_x, \Delta_y, \Delta_z\right) + \epsilon \tag{10}$$

Where the vector $\Delta_x, \Delta_y$, and $\Delta_z$ are the voxel dimensions and $\epsilon = 0.05\ mm$ for the employed SMV implementation. This was repeated until the index $i$ reached the value $i^*$ for which $i^* = i_{max}$ or for which the ratio of the size of the mask $M_{b,v}^{i^*}(\boldsymbol{r})$ and that of the original mask was below a given fraction $\alpha$. The final map reconstructed from the filtered field $b_{mSMV}(\boldsymbol{r})$ is then obtained as:

$$b_{mSMV}(\boldsymbol{r}) = M b_L^{i^*}(\boldsymbol{r}) \tag{11}$$

This method is illustrated in Figure 1.

The mSMV-filtered dipole field (Eq. 11) inversion, MEDI-mSMV, consists of solving[15,34]

$$\chi^* = \underset{\chi}{\operatorname{argmin}} \left\| W\left(e^{i b_{mSMV}} - e^{i(\delta - S_{r_1}) d \otimes \chi}\right) \right\|_2^2 + \lambda_1 \| M_G \nabla \chi \|_1 + \lambda_2 \| M_{CSF}(\chi - \bar{\chi}_{CSF}) \|_2^2 \tag{12}$$

where $W$ is the structural weighting mask, $d$ is the dipole kernel, $M_G$ the edge mask, $M_{CSF}$ is the CSF mask, and $\bar{\chi}_{CSF}$ is the mean CSF susceptibility.[14] Here, MEDI-mSMV is compared with the SMV-filtered dipole field inversion, MEDI-SMV:

$$\chi^* = \underset{\chi}{\operatorname{argmin}} \left\| W \left( e^{iM_{r_1} b_L^0} - e^{i(\delta - S_{r_1}) d \otimes \chi} \right) \right\|_2^2 + \lambda_1 \| M_G \nabla \chi \|_1 + \lambda_2 \| M_{CSF} (\chi - \bar{\chi}_{CSF}) \|_2^2 \tag{13}$$

and unfiltered dipole field inversion, MEDI:

$$\chi^* = \underset{\chi}{\operatorname{argmin}} \left\| W \left( e^{i\hat{b}_L} - e^{i(d \otimes \chi)} \right) \right\|_2^2 + \lambda_1 \| M_G \nabla \chi \|_1 + \lambda_2 \| M_{CSF} (\chi - \bar{\chi}_{CSF}) \|_2^2 \tag{14}$$

Where global CSF regularization was used.[15] The performance of the algorithms in Eq. 12 and Eq. 14 is also compared using background fields obtained using VSHARP[17] and LBV.[21]

## METHODS

### Numerical brain

#### Simulation

All studies were approved by the Weill Cornell Institutional Review Board. The true brain susceptibility was assumed to be a piecewise constant model obtained from the MEDI toolbox sample data[35] with an included brain mask and voxel size = $0.9375 \times 0.9375 \times 1.5\ mm^3$. The background air susceptibility $\chi_B$ was assumed to be $9.4\ ppm$.[36] The field was computed by convolving the susceptibility $\chi$ within and outside the brain with the dipole kernel. Multi-gradient echo signal was modeled as $S(t_j) = Ae^{-2\pi i b f_c TE_j}$ with $A$ simulated magnitude signal between 0 and 1, relative difference field $b$, central frequency $f_c = \gamma B_0$ where $B_0 = 3T$ and echo time $TE_j$. The simulated signal consisted of 11 echoes with spacing $\Delta TE = 2.6\ ms$ and additive complex Gaussian noise $\mathcal{N}(0,1/SNR)$, with $SNR = 50$.[37] These methods were repeated at a high SNR simulation, $SNR = 500$.

#### Reconstruction

The $R_2^*$ map was computed using auto-regression on linear operations (ARLO),[38] phase was unwrapped using rapid opensource minimum spanning tree algorithm (ROMEO)[39] and background field was removed using PDF,[30] or VSHARP.[17] The following methods were compared: PDF+MEDI-mSMV, PDF+MEDI-SMV (SMV-filtered dipole inversion), PDF+MEDI, and VSHARP+MEDI. MEDI-mSMV parameters were $r_1 = 5\ mm$, $t_{min} = 0.3\ Hz$, $i_{max} = 5$, and $\alpha = 10^{-6}$, MEDI-SMV parameters were $r_1 = 5\ mm$, and VSHARP[17] parameters were $r_1 = 5\ mm$ and $r_2 = 1\ mm$ with 5 layers using the SEPIA-STI implementation.[40] Kernel radii were selected from the published range[41] for each method by maximizing correlation via linear regression between the resulting tissue field with the ground truth tissue field. MEDI[1] was performed using regularization parameters $\lambda_1 = 100$ and $\lambda_2 = 100$.

### In vivo brains

Healthy subjects

This was a retrospective image analysis study approved by the Weill Cornell Institutional Review Board. On Siemens $3T$ Prisma, eleven healthy subjects underwent monopolar 3D mGRE with acquisition parameters: $FOV = 25.6\ cm$, phase $FOV$ factor $= 0.813$, $TR = 41\ ms$, $TE_1 = 2.2\ ms$, $\Delta TE = 3.25\ ms$, 7 echoes, voxel size = $1 \times 1 \times 2\ mm^3$ (interpolated to $1\ mm^3$ isotropic), $RBW = 260\ Hz/pixel$, flip angle $\alpha_{FA} = 15°$, slice partial Fourier factor = 0.875, GRAPPA acceleration factor $R = 2$, acquisition time ~6 $min$. For segmentation, whole-brain axial $T_1w$ scans were acquired using Magnetization Prepared Rapid Acquisition Gradient Echo (MPRAGE) protocol with field of view $(FOV) = 25.6\ cm$, phase $FOV$ factor = 1, repetition time $(TR) = 2300\ ms$, echo time $(TE) = 2.26\ ms$, voxel size = $1\ mm^3$ isotropic, readout bandwidth $(RBW) = 200\ Hz/pixel$, flip angle $\alpha_{FA} = 8°$, GRAPPA acceleration factor $R = 2$, acquisition time ~ 5 $min$. Tissue fields were obtained using LBV (single boundary layer peel, full multigrid parameters $N_1 = 30$, $N_2 = 100$, $N_3 = 100$), VSHARP and PDF. QSM was reconstructed using LBV+MEDI, LBV+MEDI-mSMV, VSHARP+MEDI, VSHARP+MEDI-mSMV, PDF+MEDI, PDF+MEDI-SMV, and PDF+MEDI-mSMV with the same parameters as in the “Numerical brain” section, except for the regularization parameters $\lambda_1 = 1000$ and $\lambda_2 = 100$. CSF masking was based on a background field corrected $R_2^*$ map.[42]

Registration of $T_1w$ image to mGRE magnitude was performed with FMRIB's Linear Image Registration Tool (FLIRT).[43] Gray matter masks and subcortical ROI masks were obtained from FreeSurfer atlas-based parcellation and segmentation.[44-47] The mSMV processing time was approximately 30 $s$ using an Intel i7 2.5 GHz 8-core processor.

Clinical scoring

This was a retrospective image analysis study approved by the Weill Cornell Institutional Review Board. Acquired mGRE images from forty-three patients undergoing routine clinical brain MRI across 11 scanners at Weill Cornell Medicine with field strength $B_0 = 1.5\,T - 3\,T$, slice thickness $= 1\,mm - 3\,mm$, matrix size $= 240 \times 320 - 512 \times 512$, initial $TE = 2.9\,ms - 6.7\,ms$, and final $TE = 23.5\,ms - 49\,ms$. Given increased artifacts accompanying large susceptibility sources,[12] three intracranial hemorrhage cases were included where erosion obscured some or all of the hemorrhage volume. QSMs were reconstructed with PDF+MEDI, PDF+MEDI-SMV, PDF+MEDI-mSMV, VSHARP+MEDI, and VSHARP+MEDI-mSMV using the same parameters as in the "Healthy subjects" section. Image quality related to 1) cortical gray matter, 2) veins, 3) relevant pathology 4) brain stem (to measure the clinical impact of erosion), and 5) overall shadow artifact for each method and each patient was scored by a neuroradiologist with 35 years of experience. Pathology included microhemorrhages, meningiomas, brain metastasis, cerebral infarct, pineal tumor, venous anomaly, cavernous malformation, and multiple sclerosis (MS) lesions. Each category was rated as: 1 – poor, 2 – diagnostic, 3 – good, 4 – excellent.

Multiple sclerosis patients

This was a retrospective image analysis study approved by the Weill Cornell Institutional Review Board. Fifty MS patients (containing 907 lesions) were scanned using 3D axial mGRE voxel size = $0.75 \times 0.75 \times 3\ mm^3$, acquisition matrix size = $320 \times 260 \times 50\text{–}60$, flip angle $\alpha_{FA} = 15°$, initial $TE = 6.3\ ms$, repetition time $TR = 49\ ms$, number of echoes of 10, echo spacing $\Delta TE = 4.1\ ms$ and a scan time of 4.3 $min$. QSM was reconstructed using the same parameters as in “Healthy subjects”. PDF+MEDI-SMV and PDF+MEDI-mSMV as well as VSHARP+MEDI and VSHARP+MEDI-mSMV were compared using mean susceptibility of each segmented lesion. Lesion masks were obtained from automatic segmentation of the FLAIR image, edited when necessary, by an experienced neuroradiologist. The FLAIR lesion mask was registered to QSM using FLIRT to obtain mean susceptibility.

**Statistical analysis**

Numerical brain

Shadow artifact was quantified[48] as susceptibility variance within the gray matter mask and compared using a Wilcoxon signed-rank test with significance level set at $p < 0.01$. QSM accuracy was measured by correlation and slope obtained by linear regression between the reconstructions across major subcortical gray matter ROIs (globus pallidus, putamen, thalamus, and caudate nucleus). Agreement between different in vivo reconstructions was assessed with Bland-Altman plots.

In vivo data

The analysis for the numerical brain was repeated for in vivo data. Additionally, clinical scores between different methods were compared using a Wilcoxon signed-rank test with significance level set at $p < 0.05$. Susceptibility values were compared across all lesions using linear regression.

## RESULTS

### Numerical brain

Cortical gray matter variance was lower in PDF+MEDI-mSMV ($4.20 \times 10^{-4}\ ppm^2$) and VSHARP+MEDI-mSMV ($7.71 \times 10^{-5}\ ppm^2$) compared to PDF+MEDI ($7.77 \times 10^{-4}\ ppm^2$) and VSHARP+MEDI ($6.26 \times 10^{-4}\ ppm^2$). Results for linear regression between the results of each method and the ground truth ROIs were: $r = 0.998,\ slope = 1.14$ for PDF+MEDI-mSMV, $r = 0.995$, $slope = 1.04$, for VSHARP+MEDI-mSMV, $r = 0.997,\ slope = 1.05$ for PDF+MEDI and $r = 0.998,\ slope = 1.17$ for VSHARP+MEDI. At high SNR (500), the slopes among these methods varied from 0.98 (PDF+MEDI-mSMV) to 1.04 (PDF+MEDI). Supplementary Materials contains images in three orthogonal planes (Figure S1) and the high SNR simulation results.

### Healthy subjects

Healthy subject QSMs are compared for various background field removal methods in Figure 2 and Figure 3. Lower cortical gray matter variance representing improved shadow scores is demonstrated with the use of mSMV in Figure S2. The median shadow score in VSHARP-MEDI is $0.0019\ ppm^2$ and $0.0014\ ppm^2$ in VSHARP-MEDI+mSMV, $p < 0.01$. In LBV-MEDI, the median shadow score is $0.0038\ ppm^2$ and $0.0015\ ppm^2$ in LBV-MEDI+mSMV, $p < 0.01$. The median shadow score in PDF-MEDI is $0.0039\ ppm^2$ and $0.0019\ ppm^2$ in PDF-MEDI+mSMV, $p < 0.01$.

Bland-Altman analysis (Figure S3) revealed a bias and limits of agreement of $0.013\ ppm$ and $[-0.008\ ppm, 0.036\ ppm]$ between PDF+MEDI-mSMV and PDF+MEDI, $1.5 \times 10^{-3}\ ppm$ and $[-0.005\ ppm, 0.005\ ppm]$ between PDF+MEDI-mSMV and PDF+MEDI-SMV, and $0.004\ ppm$ and $[-0.003\ ppm, 0.01\ ppm]$ between PDF+MEDI-mSMV and VSHARP+MEDI. Regression results (Figure S4) between PDF+MEDI-mSMV and each method were: $r = 0.97,\ slope = 0.95$ for PDF+MEDI, $r = 0.99,\ slope = 1.01$ for PDF+MEDI-SMV, and $r = 0.99,\ slope = 0.94$ for VSHARP+MEDI.

**Clinical scoring**

Figure 4 compares image scores across categories for PDF+MEDI, PDF+MEDI-SMV, PDF+MEDI-mSMV, VSHARP+MEDI, and VSHARP+MEDI-mSMV. The findings indicate higher image scores ($p < 0.01$) in cortical gray matter and vein depiction with PDF+MEDI-mSMV compared to PDF+MEDI and PDF+MEDI-SMV, higher scores ($p < 0.05$) in pathology depiction (pineal tumor, cerebral and thalamic microbleeds, and lesions) due to full brain coverage compared to PDF+MEDI-SMV and higher scores ($p < 0.01$) in shadow reduction compared to PDF+MEDI and VSHARP+MEDI as well as brain stem image quality compared to PDF+MEDI. Similarly, VSHARP+MEDI-mSMV has higher image score ($p < 0.01$) for shadow compared to VSHARP+MEDI and PDF+MEDI-mSMV. Clinical feedback indicates PDF+MEDI-mSMV images improve diagnostic value by adding coverage of cortical gray matter and veins (especially the superior sagittal sinus) for better depiction of cortical microbleeds and cortical superficial siderosis. Shadow artifacts and brain stem image quality were comparable to PDF+MEDI-SMV. Figure 5 shows cases with a hemorrhage near the boundary of the brain. While PDF+MEDI can visualize these lesions, there are shadow artifacts obscuring the brain. PDF+MEDI-SMV suppress these artifacts but also remove or obscure the hemorrhages due to brain erosion. In comparison, PDF+MEDI-mSMV preserves these hemorrhages while providing effective shadow reduction similar to PDF+MEDI-SMV.

**Multiple sclerosis patients**

Strong correlation $r = 0.98$ and $slope = 1.06$ was found in the lesion susceptibility linear fit between PDF+MEDI-SMV and PDF+MEDI-mSMV. The limits of agreement were $[-0.004\ ppm, 0.005\ ppm]$ with a bias of $3.3 \times 10^{-4}\ ppm$. Similar correlation and agreement ($r = 0.98$ and $slope = 1.05$, with limits $[-0.005\ ppm, 0.005\ ppm]$ and bias $2.7 \times 10^{-4}\ ppm$ was found between VSHARP+MEDI and VSHARP+MEDI-mSMV (Figure S6).

## DISCUSSION

The results show the feasibility of filtering large field magnitude values near the border based on the maximum spherical mean value (mSMV) property of the harmonic background field without eroding the brain mask in QSM. Reduced shadow is demonstrated in a numerical simulation and in healthy subjects, measured by the gray matter variance and radiological scoring. The mSMV method is robust against a variety of scan parameters encountered in patient data and improves QSM visualization of multiple sclerosis lesions and hemorrhages near the brain boundary.

The proposed mSMV method uses the maximum value property of a harmonic function to identify border voxels that have high field magnitude values and likely contain residual background field. Since the mask is not eroded as with other methods, dipole inversion using a maximum a posteriori estimator (MEDI-$L_1$) allows the entire brain volume to be preserved.

Removing residual background field via improved background field removal techniques reduces shadows, but each background field removal technique is fundamentally limited by assumptions made. For example, while the number of PDF iterations can be increased to remove residual background field to reduce some shadow artifacts, mSMV filtering still improves the QSM by decreasing shadows further still (Figure S7). The PDF background field removal method was selected due to its lack of erosion, but mSMV improves reconstructions from other background field removal techniques like VSHARP and LBV as shown in Figure 3, Figure 4 and Figures S1-S6. As demonstrated in the numerical simulations, healthy subjects, and/or clinical scoring results, VSHARP+MEDI-mSMV and LBV+MEDI-mSMV reduce shadow compared to PDF+MEDI-mSMV at the cost of 1 $mm$ (VSHARP) or 1 layer (LBV) erosion of brain tissue.

There are several limitations to this work. 1) Given the characteristic brain erosion unique to each algorithm, the scoring was not blind. 2) The minimal radius $r_2$ generates a kernel $\kappa(r_2)$ that approximates a sphere and is susceptible to discretization error. 3) The mSMV algorithm vessel mask relies on the $R_2^*$ acquisition to differentiate high tissue field values from residual background field, limiting applications to multi-echo gradient echo sequences. 4) Error in $R_2^*$ from noisy regions such as around temporal bones and nasal cavities may reduce the quality of this vessel mask. The vessels and other high local tissue field features such as microbleeds and hemorrhages are approximated by ellipsoids in the Frangi filter and omitted from the residual background field estimation to preserve their detail. This omission may lead to artifacts in the final QSM reconstruction. In the future, this may be avoided if separate SMV-filtering was applied within the Frangi filter-derived vessel mask. 5) The simulation reconstruction slopes for PDF+MEDI-SMV and PDF+MEDI-mSMV varied with SNR, suggesting susceptibility values are somewhat lower than for PDF, the exact cause of this remains to be elucidated. 6) mSMV application here is limited to the brain, which is important for studying many neurological diseases including Alzheimer's disease,[49] Parkinson's disease,[50] amyotrophic lateral sclerosis[51] and multiple sclerosis,[52] mSMV may be extended to the body outside the brain[53] with the consideration of chemical shift of fat.[54]

## CONCLUSION

In this study, the maximum Spherical Mean Value property of harmonic functions was used to identify and remove residual background field. The proposed mSMV-filtering for dipole inversion reduces shadows similar to SMV-filtered dipole inversion while preserving the brain mask, as demonstrated in numerical simulations, healthy subjects, and various pathologies in patients.

## FIGURES

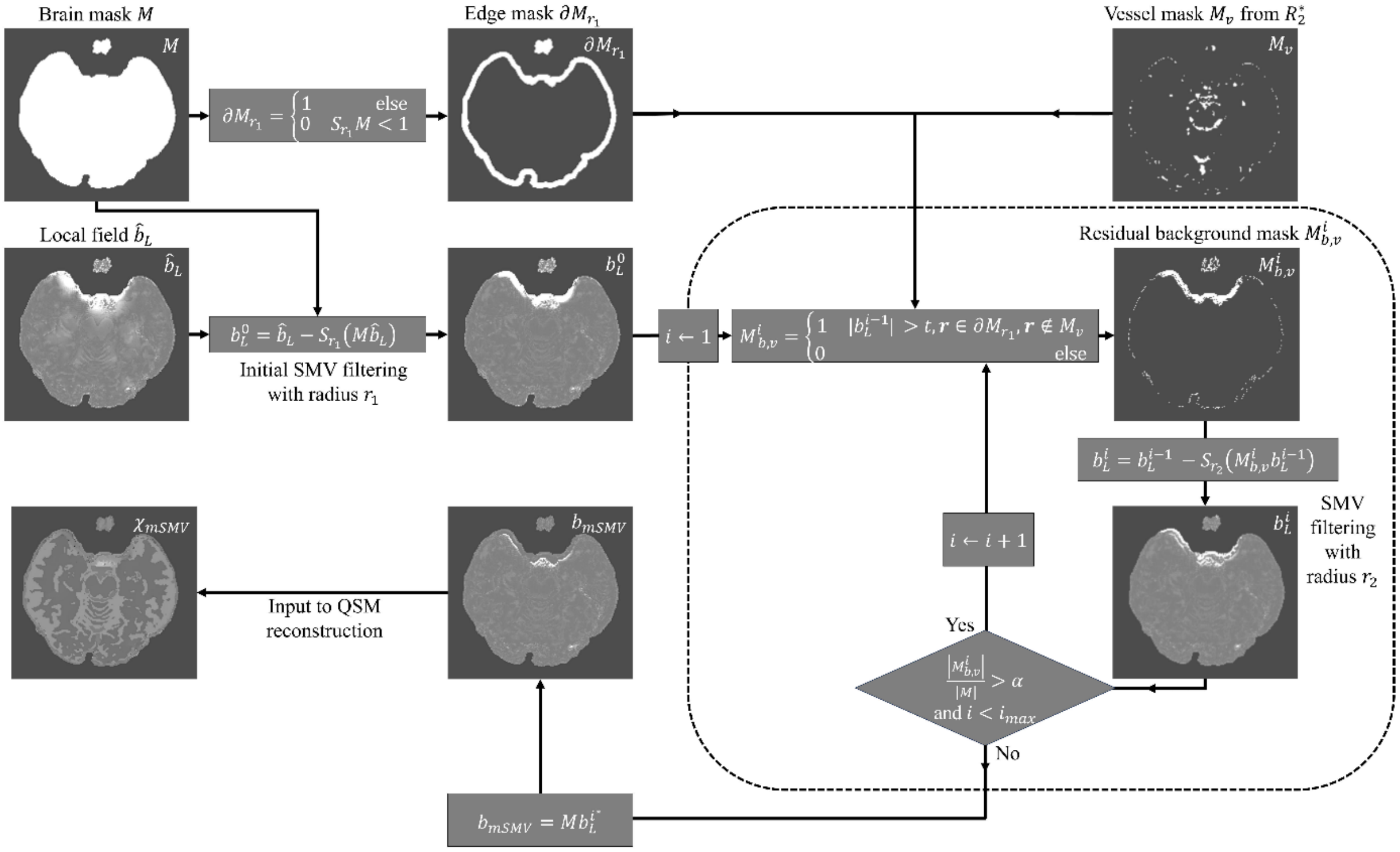

Figure 1. The mSMV algorithm first applies SMV filtering ($r_1 = 5mm$) to the whole brain input (within brain mask $M$) local tissue field $\hat{b}_L(r)$ after PDF background field removal (Eq. 4). A threshold $t$ is computed using Eq. 7. A residual background field mask $M_{b,v}^i(r)$ is constructed using this threshold, the edge mask $\partial M_{r_1}(r)$ and the vessel mask $M_v(r)$ (Eq. 8). This mask is updated at each iteration indicated by the dotted region, and voxels falling in $b_L^{i-1}(r)$ are filtered (Eq. 9) with the minimal radius kernel $r_2$ (Eq. 10). After each iteration, the local tissue field is denoted by $b_L^i(r)$. This was repeated until the index $i$ reached the value $i^*$ which satisfied $i^* = i_{max}$ or when the size of the mask $M_{b,v}^{i^*}(r)$ was a given fraction $\alpha$, relative to the size of the original mask $M$. Finally, $b_{mSMV}(r)$ (Eq. 11) is reconstructed to generate the QSM, $\chi_{mSMV}$.

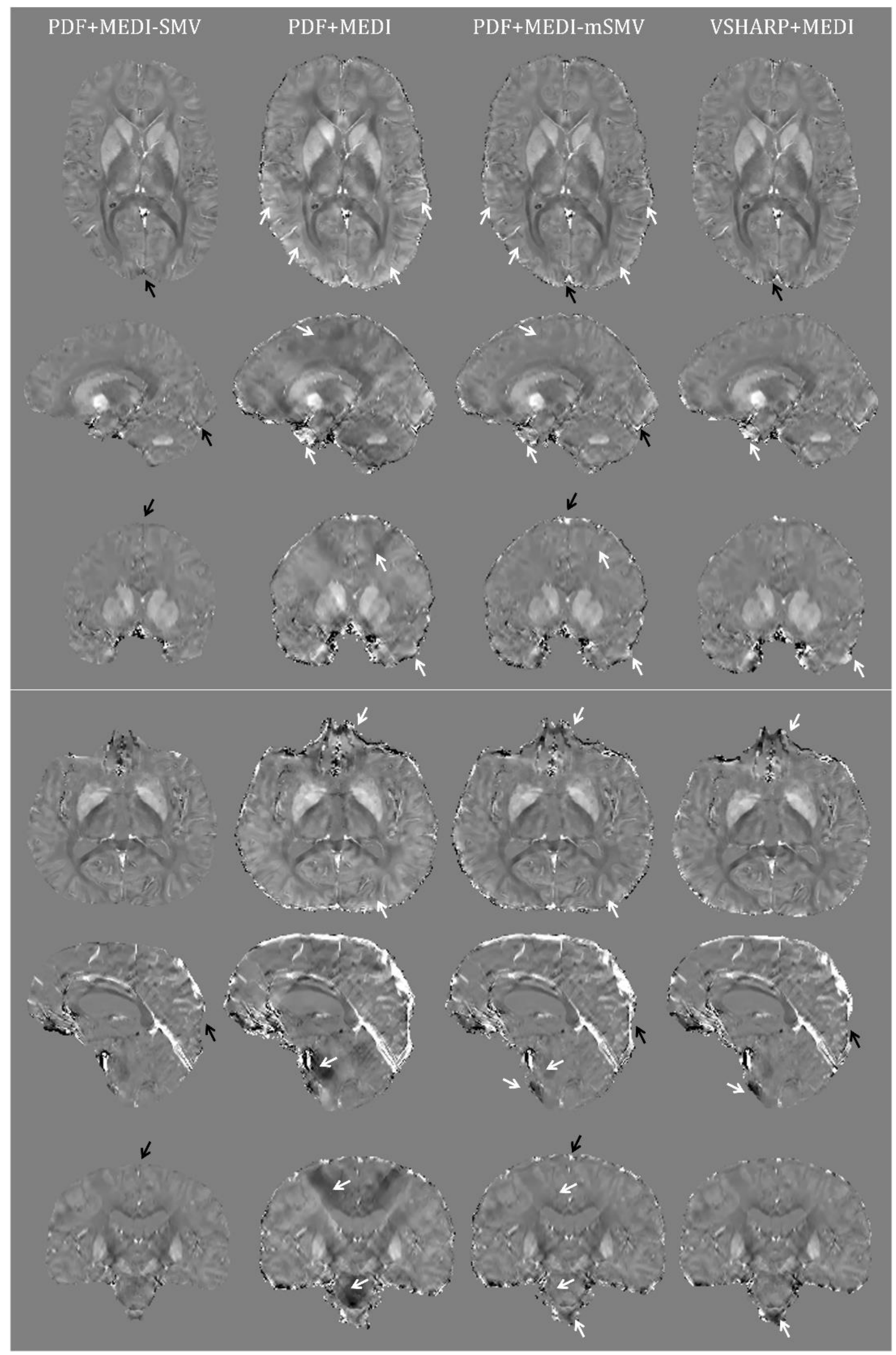

Figure 2. Comparison of QSM reconstructions in healthy subjects obtained for various background removal and dipole inversion methods. White arrows indicate areas of bright and dark shadow reduction achieved without erosion (black arrows) by mSMV.

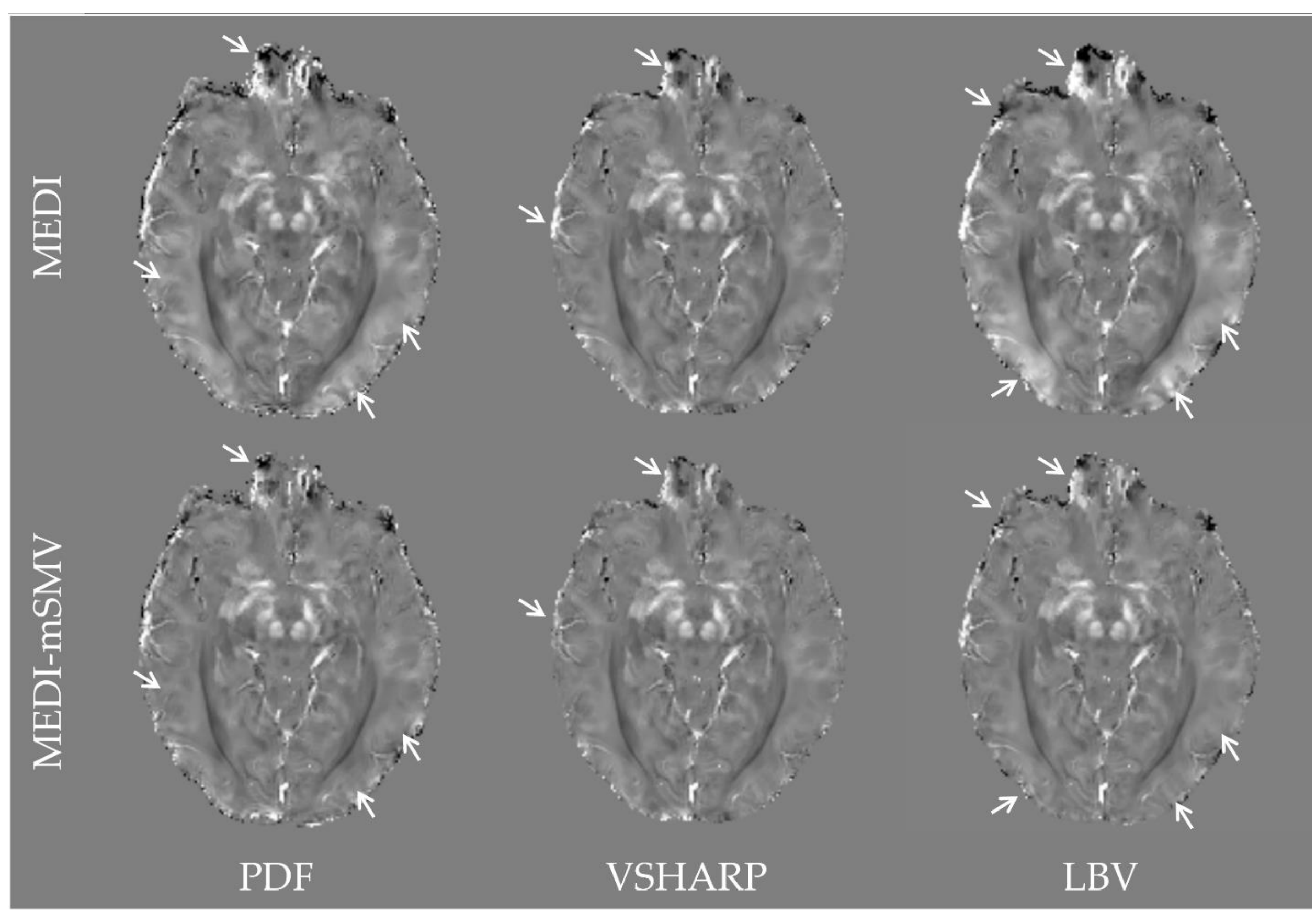

Figure 3. Comparison of QSM reconstructions for various background field removal techniques. Arrows indicate regions where mSMV reduces bright and dark shadow artifact in the healthy subject.

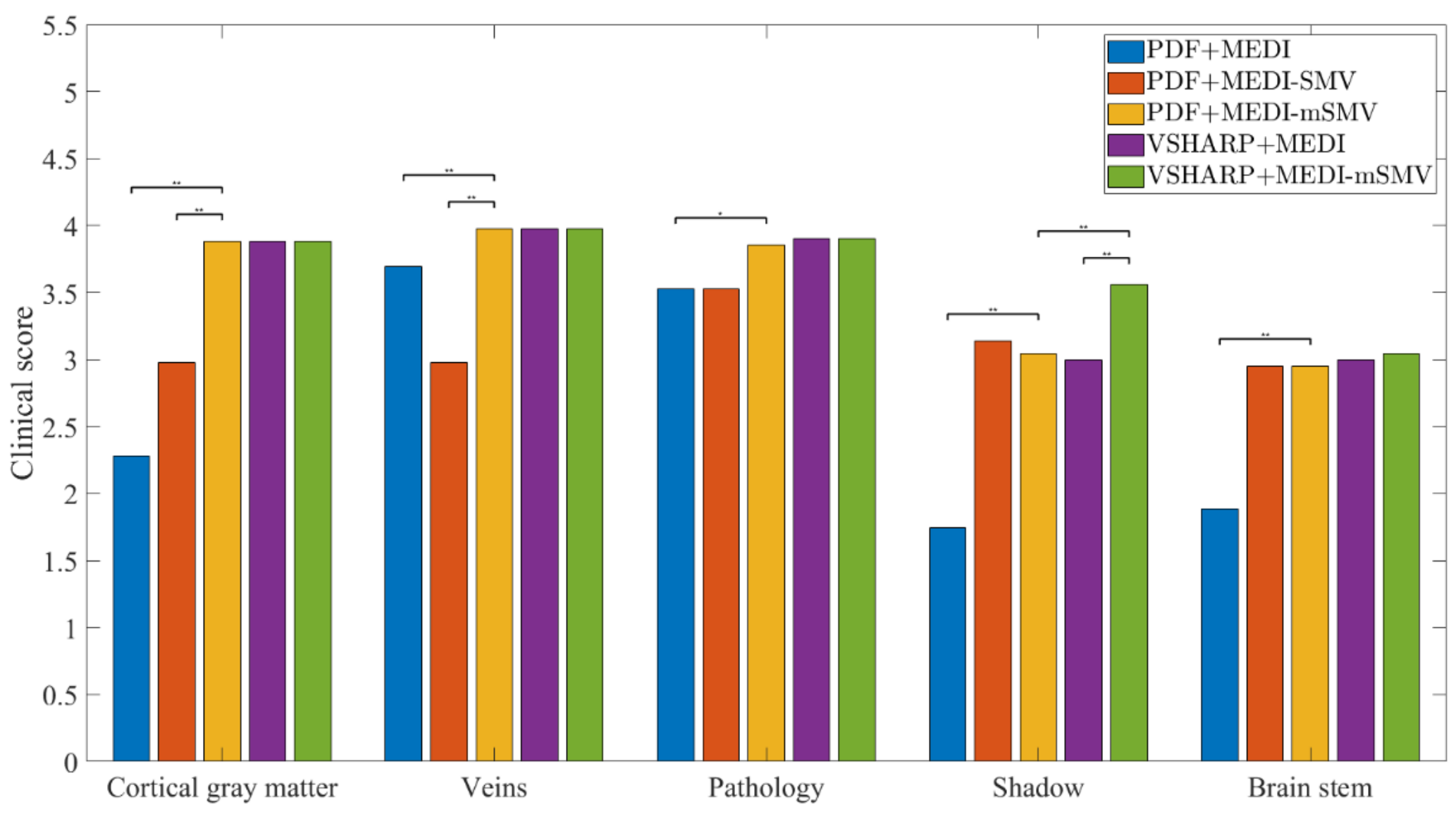

Figure 4. Clinical scoring across cortical gray matter, veins, relevant pathology, shadow artifacts, and brain stem for PDF+MEDI, PDF+MEDI-SMV, and PDF+MEDI-mSMV. Improvements are noted in each category as compared to existing methods, * indicates $p < 0.05$ and ** indicates $p < 0.01$.

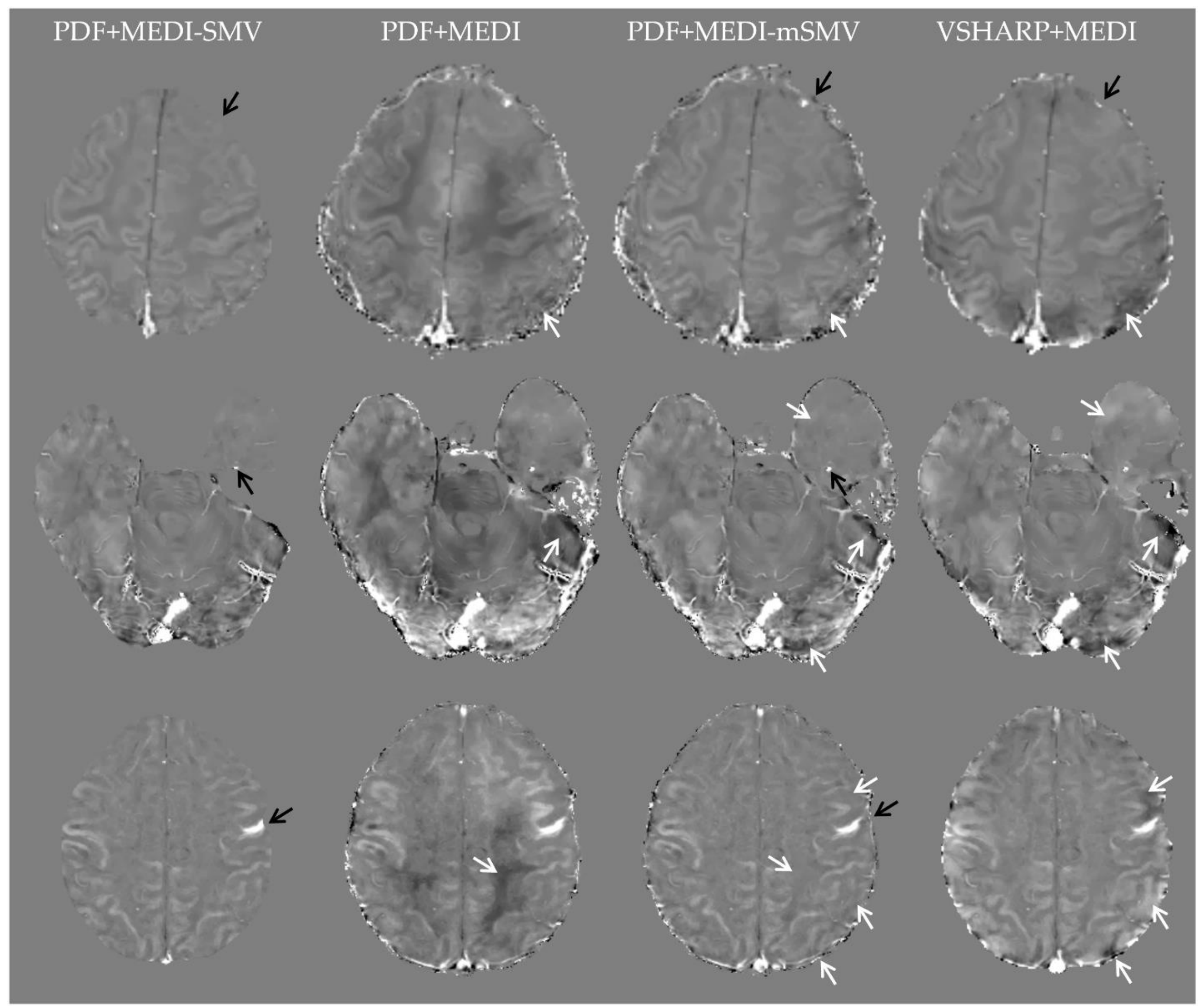

Figure 5. Comparison of PDF+MEDI-SMV, PDF+MEDI, PDF+MEDI-SMV, and VSHARP+MEDI in three patients with intracranial microhemorrhages (first and second rows) and an intracranial hemorrhage (third row). The bleed in the first row is preserved by mSMV but eroded or obscured in other methods. White arrows indicate areas of bright and dark shadow reduction achieved without erosion (black arrows).

# ACKNOWLEDGMENTS

## Funding sources

This research was partly supported by grants from the National Institutes of Health (R01 NS095562, R01 NS105144, R01 NS090464, R01 HL151686, R01 NS123576, and S10 OD021782) and the National MS Society (RR-1602-07671).

## Conflict of interest

P.S. and Y.W. are co-inventors on QSM-related patents owned by Cornell University and have ownership shares in MedImageMetric LLC. T.N. is a consultant of MedImageMetric. The remaining authors have no conflicts of interest to report.

## DATA AVAILABILITY

Data and software can be found at https://github.com/agr78/mSMV upon manuscript publication.

## REFERENCES

1. De Rochefort L, Liu T, Kressler B, et al. Quantitative susceptibility map reconstruction from MR phase data using bayesian regularization: Validation and application to brain imaging. *Magnetic Resonance in Medicine*. 2010;63(1):194-206. doi:10.1002/mrm.22187
2. Marques JP, Bowtell R. Application of a Fourier-based method for rapid calculation of field inhomogeneity due to spatial variation of magnetic susceptibility. *Concepts in Magnetic Resonance Part B: Magnetic Resonance Engineering*. 2005;25B(1):65-78. doi:10.1002/cmr.b.20034
3. Li L, Leigh JS. Quantifying arbitrary magnetic susceptibility distributions with MR. *Magnetic Resonance in Medicine*. 2004;51(5):1077-1082. doi:10.1002/mrm.20054
4. Wang Y, Liu T. Quantitative susceptibility mapping (QSM): Decoding<scp>MRI</scp>data for a tissue magnetic biomarker. *Magnetic Resonance in Medicine*. 2015;73(1):82-101. doi:10.1002/mrm.25358
5. Wharton S, Schäfer A, Bowtell R. Susceptibility mapping in the human brain using threshold-based k-space division. *Magnetic Resonance in Medicine*. 2010;63(5):1292-1304. doi:10.1002/mrm.22334
6. Choi JK, Park HS, Wang S, Wang Y, Seo JK. Inverse Problem in Quantitative Susceptibility Mapping. *SIAM Journal on Imaging Sciences*. 2014;7(3):1669-1689. doi:10.1137/140957433
7. Youngwook Kee ZL, Liangdong Zhou, Alexey Dimov, Junghun Cho, Ludovic de Rochefort, Jin Keun Seo, and Yi Wang. Quantitative Susceptibility Mapping (QSM) Algorithms: Mathematical Rationale and Computational Implementations. *IEEE Transactions on Biomedical Engineering*. 2017;64(11):2531-2545. doi:10.1109/tbme.2017.2749298
8. Zhou L, Choi JK, Kee Y, Wang Y, Seo JK. Dipole Incompatibility Related Artifacts in Quantitative Susceptibility Mapping. *arXiv: Medical Physics*. 2017;
9. Liu T, Liu J, De Rochefort L, et al. Morphology enabled dipole inversion (MEDI) from a single-angle acquisition: Comparison with COSMOS in human brain imaging. *Magnetic Resonance in Medicine*. 2011;66(3):777-783. doi:10.1002/mrm.22816
10. Milovic C, Bilgic B, Zhao B, Acosta-Cabronero J, Tejos C. Fast nonlinear susceptibility inversion with variational regularization. *Magnetic Resonance in Medicine*. 2018;80(2):814-821. doi:10.1002/mrm.27073
11. Acosta-Cabronero J, Milovic C, Mattern H, Tejos C, Speck O, Callaghan MF. A robust multi-scale approach to quantitative susceptibility mapping. *Neuroimage*. Dec 2018;183:7-24. doi:10.1016/j.neuroimage.2018.07.065
12. Wei H, Dibb R, Zhou Y, et al. Streaking artifact reduction for quantitative susceptibility mapping of sources with large dynamic range. *NMR in Biomedicine*. 2015;28(10):1294-1303. doi:10.1002/nbm.3383
13. Stewart AW, Robinson SD, O'Brien K, et al. QSMxT: Robust masking and artifact reduction for quantitative susceptibility mapping. *Magnetic Resonance in Medicine*. 2022;87(3):1289-1300. doi:10.1002/mrm.29048
14. Liu Z, Spincemaille P, Yao Y, Zhang Y, Wang Y. MEDI+0: Morphology enabled dipole inversion with automatic uniform cerebrospinal fluid zero reference for quantitative

susceptibility mapping. *Magnetic Resonance in Medicine*. 2018;79(5):2795-2803. doi:10.1002/mrm.26946
15. Dimov AV, Nguyen TD, Spincemaille P, et al. Global cerebrospinal fluid as a zero-reference regularization for brain quantitative susceptibility mapping. *Journal of Neuroimaging*. 2022;32(1):141-147. doi:10.1111/jon.12923
16. Schweser F, Deistung A, Lehr BW, Reichenbach JR. Quantitative imaging of intrinsic magnetic tissue properties using MRI signal phase: An approach to in vivo brain iron metabolism? *NeuroImage*. 2011;54(4):2789-2807. doi:10.1016/j.neuroimage.2010.10.070
17. Li W, Wu B, Liu C. Quantitative susceptibility mapping of human brain reflects spatial variation in tissue composition. *NeuroImage*. 2011;55(4):1645-1656. doi:10.1016/j.neuroimage.2010.11.088
18. Sun H, Wilman AH. Background field removal using spherical mean value filtering and Tikhonov regularization. *Magnetic Resonance in Medicine*. 2014;71(3):1151-1157. doi:10.1002/mrm.24765
19. Fang J, Bao L, Li X, Van Zijl PCM, Chen Z. Background field removal for susceptibility mapping of human brain with large susceptibility variations. *Magnetic Resonance in Medicine*. 2019;81(3):2025-2037. doi:10.1002/mrm.27492
20. Wen Y, Zhou D, Liu T, Spincemaille P, Wang Y. An iterative spherical mean value method for background field removal in MRI. *Magnetic Resonance in Medicine*. 2014;72(4):1065-1071. doi:10.1002/mrm.24998
21. Zhou D, Liu T, Spincemaille P, Wang Y. Background field removal by solving the Laplacian boundary value problem. *NMR in Biomedicine*. 2014;27(3):312-319. doi:10.1002/nbm.3064
22. Schweitzer AD, Liu T, Gupta A, et al. Quantitative Susceptibility Mapping of the Motor Cortex in Amyotrophic Lateral Sclerosis and Primary Lateral Sclerosis. *American Journal of Roentgenology*. 2015;204(5):1086-1092. doi:10.2214/ajr.14.13459
23. Azuma M, Hirai T, Yamada K, et al. Lateral Asymmetry and Spatial Difference of Iron Deposition in the Substantia Nigra of Patients with Parkinson Disease Measured with Quantitative Susceptibility Mapping. *American Journal of Neuroradiology*. 2016;37(5):782-788. doi:10.3174/ajnr.a4645
24. Wang Y, Spincemaille P, Liu Z, et al. Clinical quantitative susceptibility mapping (QSM): Biometal imaging and its emerging roles in patient care. *Journal of Magnetic Resonance Imaging*. 2017;46(4):951-971. doi:10.1002/jmri.25693
25. Langkammer C, Bredies K, Poser BA, et al. Fast quantitative susceptibility mapping using 3D EPI and total generalized variation. *NeuroImage*. 2015;111:622-630. doi:10.1016/j.neuroimage.2015.02.041
26. Roberts A, Spincemaille P, Nguyen T, Wang Y. MEDI-FM: Field Map Error Guided Regularization for Shadow Reduction in Quantitative Susceptibility Mapping. presented at: International Society for Magnetic Resonance in Medicine; 2022; London, England. https://archive.ismrm.org/2022/2359.html
27. Roberts A, Spincemaille P, Nguyen T, Wang Y. MEDI-d: Downsampled Morphological Priors for Shadow Reduction in Quantitative Susceptibility Mapping. presented at: International Society for Magnetic Resonance in Medicine; 2021; Vancouver, Canada. https://cds.ismrm.org/protected/21MPresentations/abstracts/2599.html
28. Roy KK. Green's Theorem in Potential Theory. Springer Berlin Heidelberg; 307-327.

29. Rolfsen D. Harmonic Functions. University of British Columbia. Accessed 07/19, 2022. https://personal.math.ubc.ca/~rolfsen/ma317/317notes.pdf
30. Liu T, Khalidov I, De Rochefort L, et al. A novel background field removal method for MRI using projection onto dipole fields (PDF). *NMR in Biomedicine*. 2011;24(9):1129-1136. doi:10.1002/nbm.1670
31. Schweser F, Robinson SD, De Rochefort L, Li W, Bredies K. An illustrated comparison of processing methods for phase MRI and QSM: removal of background field contributions from sources outside the region of interest. *NMR in Biomedicine*. 2017;30(4):e3604. doi:10.1002/nbm.3604
32. Roberts A, Spincemaille P, Nguyen T, Wang Y. Whole Brain Spherical Mean Value Filtering for Shadow Reduction in Quantitative Susceptibility Mapping. Paper No. 2172. presented at: International Society for Magnetic Resonance in Medicine; 2023; Toronto, Canada.
33. Frangi AF, Niessen WJ, Vincken KL, Viergever MA. Multiscale vessel enhancement filtering. Springer Berlin Heidelberg; 1998:130-137.
34. Nguyen TD, Wen Y, Du J, et al. Quantitative susceptibility mapping of carotid plaques using nonlinear total field inversion: Initial experience in patients with significant carotid stenosis. *Magnetic Resonance in Medicine*. 2020;84(3):1501-1509. doi:10.1002/mrm.28227
35. Wisnieff C, Liu T, Spincemaille P, Wang S, Zhou D, Wang Y. Magnetic susceptibility anisotropy: Cylindrical symmetry from macroscopically ordered anisotropic molecules and accuracy of MRI measurements using few orientations. *NeuroImage*. 2013;70:363-376. doi:10.1016/j.neuroimage.2012.12.050
36. Schenck JF. The role of magnetic susceptibility in magnetic resonance imaging: MRI magnetic compatibility of the first and second kinds. *Med Phys*. Jun 1996;23(6):815-50. doi:10.1118/1.597854
37. Gudbjartsson H, Patz S. The Rician Distribution of Noisy MRI Data. *Magnetic Resonance in Medicine*. 1995;34(6):910-914. doi:10.1002/mrm.1910340618
38. Pei M, Nguyen TD, Thimmappa ND, et al. Algorithm for fast monoexponential fitting based on Auto-Regression on Linear Operations (ARLO) of data. *Magnetic Resonance in Medicine*. 2015;73(2):843-850. doi:10.1002/mrm.25137
39. Dymerska B, Eckstein K, Bachrata B, et al. Phase unwrapping with a rapid opensource minimum spanning tree algorithm (ROMEO). *Magnetic Resonance in Medicine*. 2021;85(4):2294-2308. doi:10.1002/mrm.28563
40. Chan K-S, Marques JP. SEPIA—Susceptibility mapping pipeline tool for phase images. *NeuroImage*. 2021;227:117611. doi:10.1016/j.neuroimage.2020.117611
41. Özbay PS, Deistung A, Feng X, Nanz D, Reichenbach JR, Schweser F. A comprehensive numerical analysis of background phase correction with V-SHARP. *NMR in Biomedicine*. 2017;30(4):e3550. doi:10.1002/nbm.3550
42. Li J, Huang W, Luo X, et al. The central vein sign in multiple sclerosis lesions: Susceptibility relaxation optimization from a routine MRI multiecho gradient echo sequence. *Journal of Neuroimaging*. 2022;32(1):48-56. doi:10.1111/jon.12938
43. Greve DN, Fischl B. Accurate and robust brain image alignment using boundary-based registration. *NeuroImage*. 2009;48(1):63-72. doi:10.1016/j.neuroimage.2009.06.060
44. Fischl B, Salat DH, Busa E, et al. Whole Brain Segmentation. *Neuron*. 2002;33(3):341-355. doi:10.1016/s0896-6273(02)00569-x

45. Segonne F, Pacheco J, Fischl B. Geometrically Accurate Topology-Correction of Cortical Surfaces Using Nonseparating Loops. *IEEE Transactions on Medical Imaging*. 2007;26(4):518-529. doi:10.1109/tmi.2006.887364

46. Fischl B, Liu A, Dale AM. Automated manifold surgery: constructing geometrically accurate and topologically correct models of the human cerebral cortex. *IEEE Transactions on Medical Imaging*. 2001;20(1):70-80. doi:10.1109/42.906426

47. Fischl B, van der Kouwe A, Destrieux C, et al. Automatically parcellating the human cerebral cortex. *Cereb Cortex*. Jan 2004;14(1):11-22. doi:10.1093/cercor/bhg087

48. Balasubramanian PS, Spincemaille P, Guo L, Huang W, Kovanlikaya I, Wang Y. Spatially Adaptive Regularization in Total Field Inversion for Quantitative Susceptibility Mapping. *iScience*. 2020;23(10):101553. doi:10.1016/j.isci.2020.101553

49. Acosta-Cabronero J, Williams GB, Cardenas-Blanco A, Arnold RJ, Lupson V, Nestor PJ. In vivo quantitative susceptibility mapping (QSM) in Alzheimer's disease. *PloS one*. 2013;8(11):e81093.

50. Azuma M, Hirai T, Yamada K, et al. Lateral Asymmetry and Spatial Difference of Iron Deposition in the Substantia Nigra of Patients with Parkinson Disease Measured with Quantitative Susceptibility Mapping. *AJNR Am J Neuroradiol*. May 2016;37(5):782-8. doi:10.3174/ajnr.A4645

51. Schweitzer AD, Liu T, Gupta A, et al. Quantitative susceptibility mapping of the motor cortex in amyotrophic lateral sclerosis and primary lateral sclerosis. *AJR Am J Roentgenol*. May 2015;204(5):1086-92. doi:10.2214/AJR.14.13459

52. Wisnieff C, Ramanan S, Olesik J, Gauthier S, Wang Y, Pitt D. Quantitative susceptibility mapping (QSM) of white matter multiple sclerosis lesions: Interpreting positive susceptibility and the presence of iron. *Magn Reson Med*. Aug 2015;74(2):564-70. doi:10.1002/mrm.25420

53. Dimov AV, Li J, Nguyen TD, et al. QSM Throughout the Body. *J Magn Reson Imaging*. Feb 7 2023;doi:10.1002/jmri.28624

54. Dimov AV, Liu T, Spincemaille P, et al. Joint estimation of chemical shift and quantitative susceptibility mapping (chemical QSM). *Magn Reson Med*. Jun 2015;73(6):2100-10. doi:10.1002/mrm.25328

## SUPPORTING INFORMATION

The following supporting information is available as part of the online article:

- Reconstruction comparisons (Figures S1-S5)

- Multiple sclerosis lesion susceptibility (Figure S6)

- Effect of increased iterations in Projection onto Dipole Fields (Figure S7)

Reconstruction comparisons (Figures S1-S5)

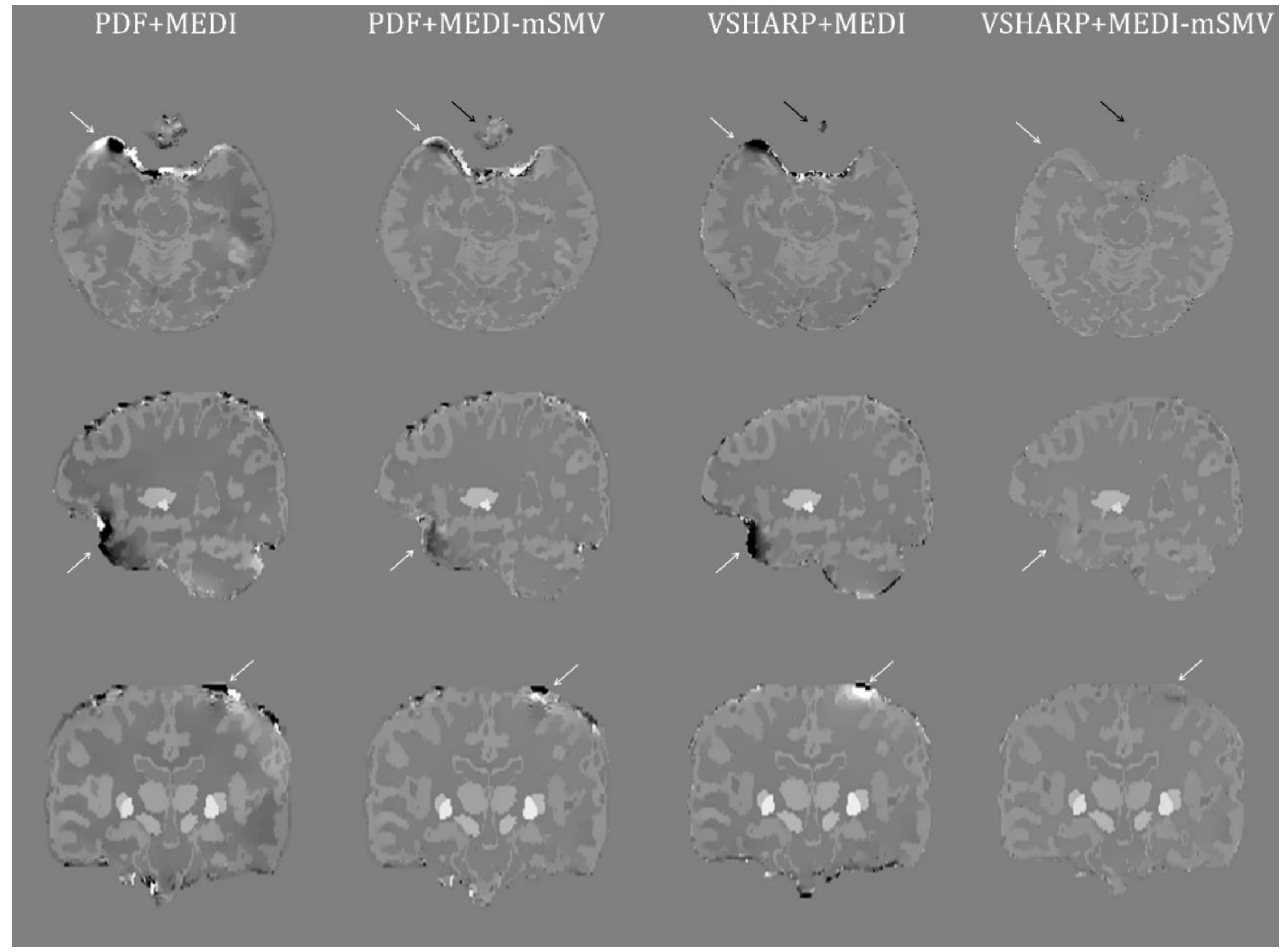

Figure S1: Comparison of numerical simulation QSM reconstructions at $SNR = 50$ obtained for various background removal and dipole inversion methods. White arrows indicate areas of bright and dark shadow reduction achieved without erosion (black arrows) by PDF+MEDI-mSMV. At $SNR = 500$, the correlations and shadow scores are $r = 0.99$, $slope = 1.04$ and $\sigma^2_{gray\ matter} = 5.88 \times 10^{-4}\ ppm^2$ for PDF+MEDI, $r = 0.99$, $slope = 0.98$ and $\sigma^2_{gray\ matter} = 1.77 \times 10^{-4}\ ppm^2$ for PDF+MEDI-SMV, $r = 0.99$, $slope = 1.00$ and $\sigma^2_{gray\ matter} = 3.92 \times 10^{-4}\ ppm^2$ for VSHARP+MEDI, $r = 0.99$, $slope = 1.02$ and $\sigma^2_{gray\ matter} = 1.41 \times 10^{-5}\ ppm^2$ for VSHARP+MEDI-mSMV, and $r = 0.99, slope = 0.98$ and $\sigma^2_{gray\ matter} = 3.18 \times 10^{-4}\ ppm^2$ for PDF+MEDI-mSMV.

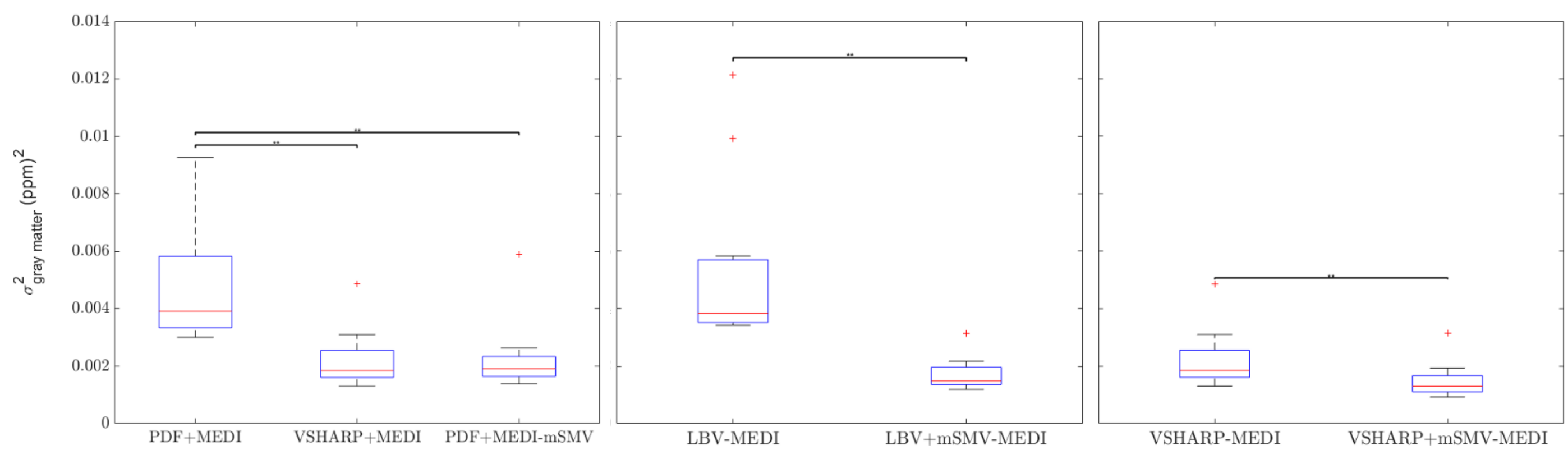

Figure S2. Application of mSMV to PDF, LBV and VSHARP background field removal techniques in healthy subjects. The median shadow score in VSHARP-MEDI is 0.0019 $ppm^2$ and 0.0014 $ppm^2$ in VSHARP-MEDI+mSMV, $p < 0.01$. In LBV-MEDI, the median shadow score is 0.0038 $ppm^2$ and 0.0015 $ppm^2$ in LBV-MEDI+mSMV, $p < 0.01$. The median shadow score in PDF-MEDI is 0.0039 $ppm^2$ and 0.0019 $ppm^2$ in PDF-MEDI+mSMV, $p < 0.01$.

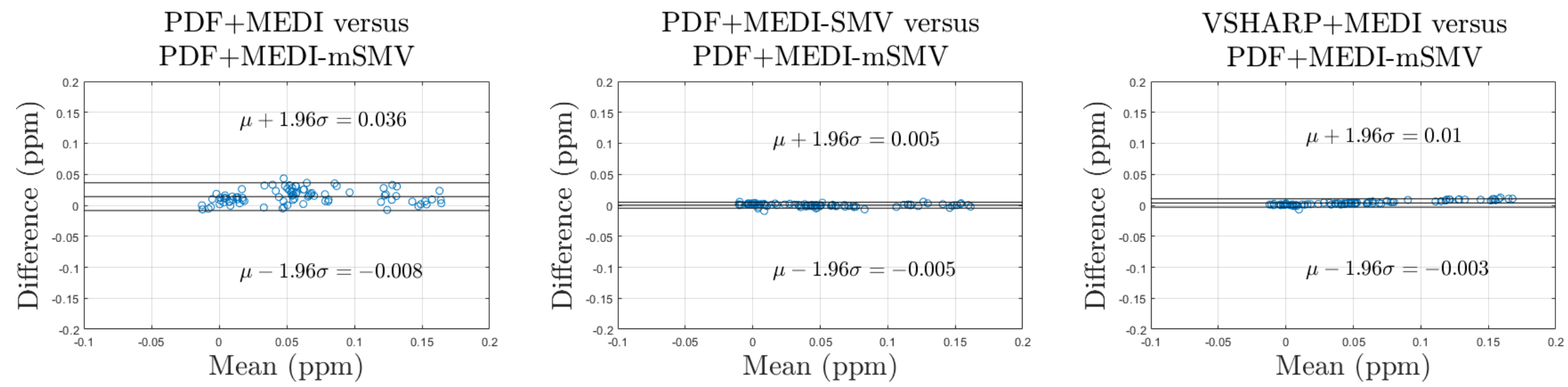

Figure S3. Bland-Altman analysis of healthy subjects reconstructions revealed a bias and limits of agreement of $0.013\ ppm$ and $[-0.008\ ppm, 0.036\ ppm]$ between PDF+MEDI-mSMV and PDF+MEDI, $1.5 \times 10^{-3}\ ppm$ and $[-0.005\ ppm, 0.005\ ppm]$ between PDF+MEDI-mSMV and PDF+MEDI-SMV, and $0.004\ ppm$ and $[-0.003\ ppm, 0.01\ ppm]$ between PDF+MEDI-mSMV and VSHARP+MEDI.

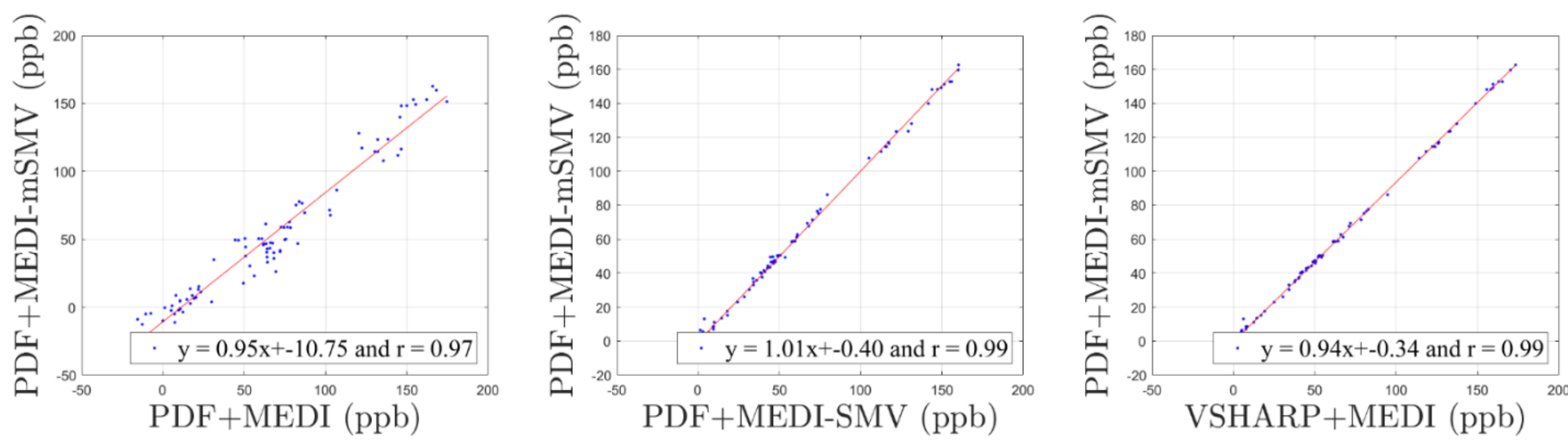

Figure S4: Subcortical gray matter fits for all healthy subjects reconstructions. The correlation and slope between PDF+MEDI-mSMV and each method was: PDF+MEDI (0.97,0.95), PDF+MEDI-SMV (0.99,1.01), VSHARP+MEDI (0.99,0.94).

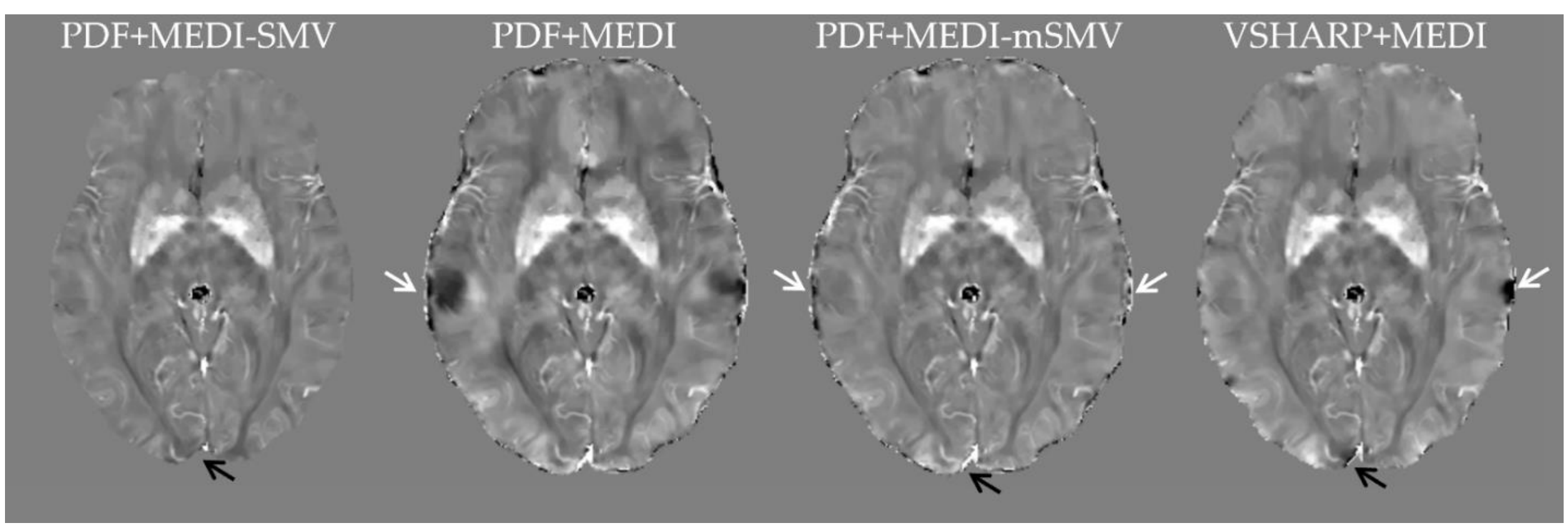

Figure S5: Shadow reduction (white arrows) without erosion (black arrow) in a 55-year-old female patient.

Multiple sclerosis lesion susceptibility (Figure S6)

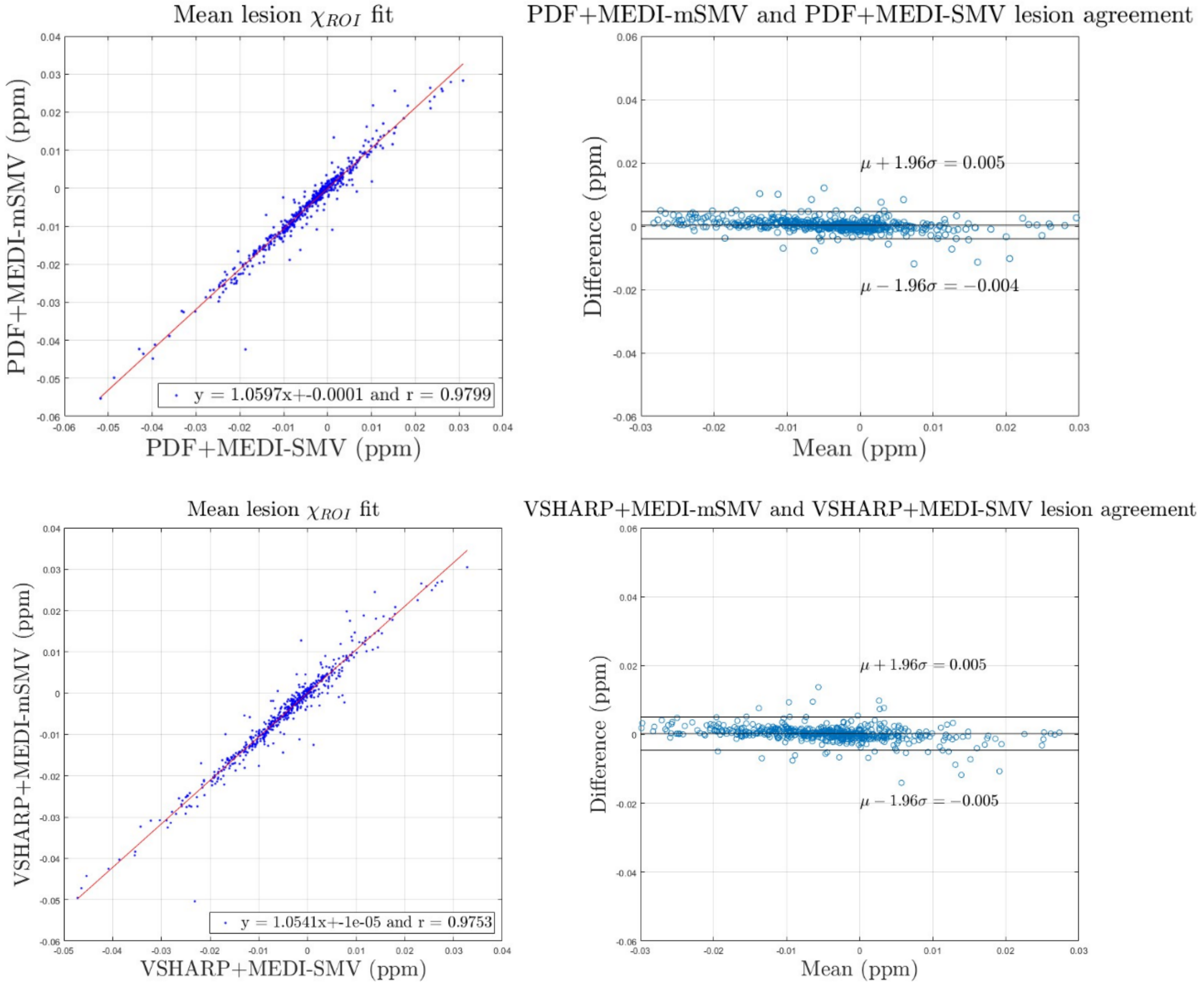

Figure S6: Fit and Bland-Altman for MS reconstructions. Strong correlation $r = 0.98$ and $slope = 1.06$ was found in the lesion susceptibility linear fit between PDF+MEDI-SMV and PDF+MEDI-mSMV. The limits of agreement were $[-0.004\ ppm, 0.005\ ppm]$ with a bias of $3.3 \times 10^{-4}\ ppm$. Similar correlation and agreement ($r = 0.98$ and $slope = 1.05$, with limits $[-0.005\ ppm, 0.005\ ppm]$ and bias $2.7 \times 10^{-4}\ ppm$ was found between VSHARP+MEDI and VSHARP+MEDI-mSMV (Figure S6).

Effect of increased iterations in Projection onto Dipole Fields (Figure S7)

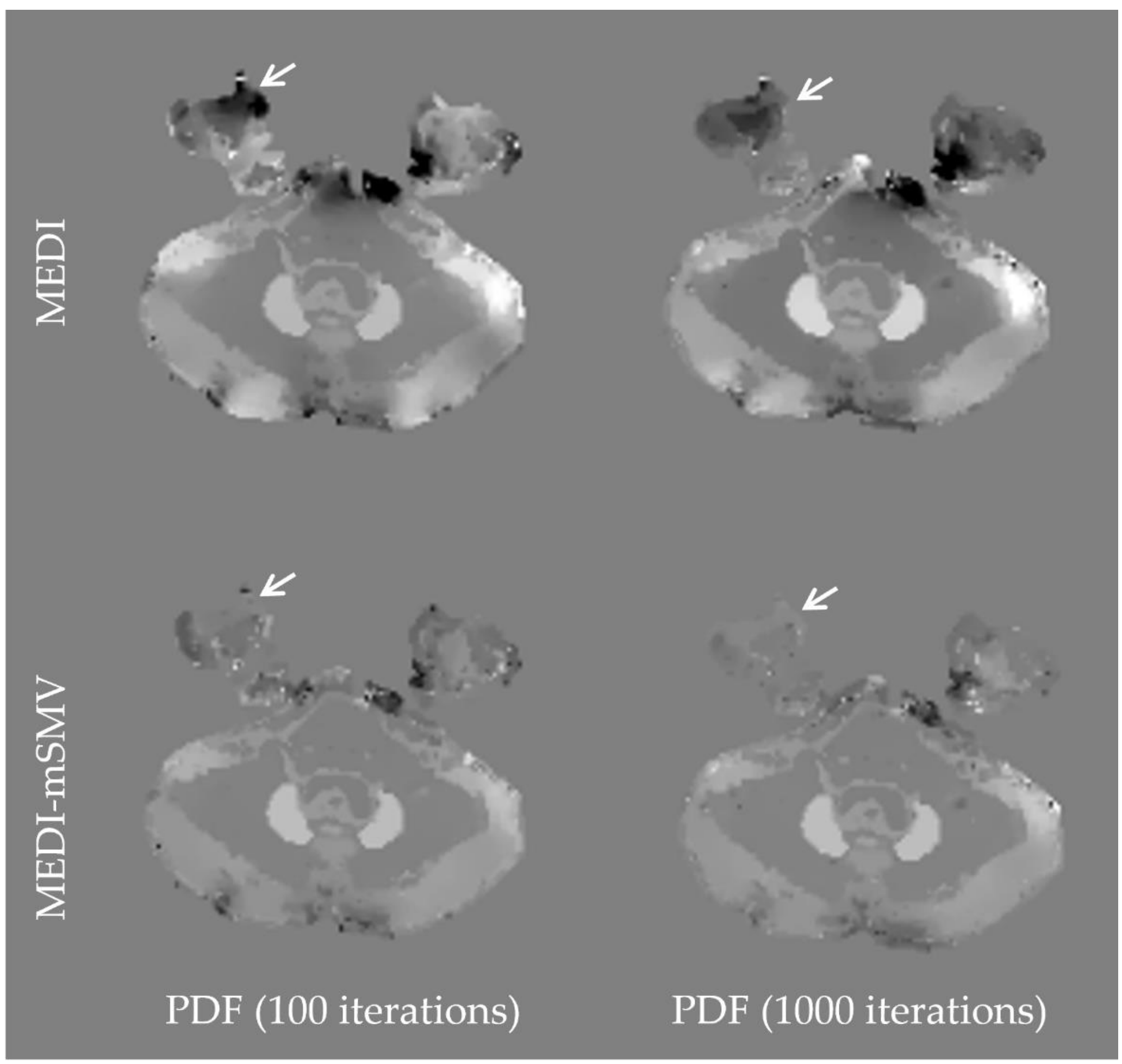

Figure S7. Improvement of numerical brain QSM after 1000 iterations of PDF background field removal.